\documentclass[%
 reprint,
showpacs,preprintnumbers,
 amsmath,amssymb, aps,pre,
floatfix,
]{revtex4-1}

\usepackage{graphicx}
\usepackage{bm}
\usepackage{subfigure}
\usepackage[usenames,dvipsnames]{color}

\begin{document}

\preprint{APS/123-QED}

\title{Stochastic mean field formulation of the dynamics of diluted neural networks}

\author{D. Angulo-Garcia}
\email{david.angulo@fi.isc.cnr.it}
\affiliation{CNR - Consiglio Nazionale delle Ricerche -
Istituto dei Sistemi Complessi, via Madonna del Piano 10,
I-50019 Sesto Fiorentino, Italy}%
\author{A. Torcini}%
\email{alessandro.torcini@cnr.it}
\affiliation{CNR - Consiglio Nazionale delle Ricerche -
Istituto dei Sistemi Complessi, via Madonna del Piano 10,
I-50019 Sesto Fiorentino, Italy}
\affiliation{INFN Sez. Firenze, via Sansone, 1 - I-50019 Sesto Fiorentino, Italy}




\date{\today}

\begin{abstract}
We consider pulse-coupled Leaky Integrate-and-Fire neural networks
with randomly distributed synaptic couplings. 
This random dilution induces fluctuations in the evolution of 
the macroscopic variables and deterministic chaos at the microscopic level.
Our main aim is to mimic the effect of the dilution as a noise source  acting on 
the dynamics of a globally coupled non-chaotic system.
Indeed, the evolution of a diluted neural network can be well approximated 
as a fully pulse coupled network, where each neuron is driven by a mean
synaptic current plus additive noise.
These terms represent the average and the fluctuations of the
synaptic currents acting on the single neurons in the diluted
system. The main microscopic and macroscopic dynamical features 
can be retrieved with this stochastic approximation. 
Furthermore, the microscopic stability  of the diluted network can be 
also reproduced, as demonstrated from the almost coincidence of the measured Lyapunov 
exponents in the deterministic and stochastic cases for an ample range
of system sizes. Our results strongly suggest that the fluctuations in the
synaptic currents are responsible for the emergence of chaos in this 
class of pulse coupled networks. 
\end{abstract}

\maketitle


\section{Introduction}

In pionieristic studies devoted to excitatory pulse-coupled 
networks of leaky integrate-and-fire (LIF) 
neurons~\cite{Abbott1993Asynchronous,vanVreeswijk1996Partial},
Abbott and van Vreeswiijk have shown that these models in a
globally coupled configuration can exhibit only two kind of evolution, both regular.
The first one, termed splay state, is associated to collective asynchronous dynamics
and the second one called partial synchronization corresponds to coherent periodic 
activity in the network. The latter regime is characterized by 
periodic oscillations in the neural activity and by quasi-periodic motions 
of the single neuron membrane potentials~\cite{vanVreeswijk1996Partial}.
The introduction of random dilution in such network, achieved by considering 
an Erd\"os-R\'{e}nyi distribution for the connectivity degrees,
induces chaoticity in the system and fluctuations in the collective activity~\cite{Olmi2010Oscillations}.
Fluctuations and chaos are due to the non equivalence of the neurons in the network.
However, for massively connected networks, where the average in-degree is proportional to
the system size, the dynamics becomes regular in the thermodynamic limit, 
recovering the evolution of the globally coupled system~\cite{Olmi2010Oscillations,Tattini2011Coherent}.
On the other hand, for sparse networks, where the in-degree value is constant
independently of the network size, the system remains chaotic for any network size~\cite{Luccioli2012PRL}.

A fundamental question  which we would like to address in this 
paper is whether the effect of the frozen network heterogeneity can be reproduced in terms 
of a homogeneous model with additive noise. In particular, 
we are interested in reproducing the chaotic behavior observed in the diluted system.
As homogeneous model we consider a fully coupled (FC) network displaying
only regular motions and we focus on the partially synchronized regime,
where the macroscopic variables are periodic~\cite{vanVreeswijk1996Partial}.
The addition of noise to the membrane potential evolution induces irregular oscillations in the 
macroscopic evolution and in the neuronal dynamics. To reproduce the dynamics of a specific 
deterministic diluted (DD) network, we employ in the stochastic model as noise amplitudes 
the ones derived from the original system. As a result, the stochastic model is able to mimic 
the main microscopic and  macroscopic features of the original diluted system and even the 
chaoticity properties of the deterministic system.  Furthermore, we are able to
mimic the dynamics of networks composed by thousands of neurons
by employing a stochastic model with only one hundred elements. 

This study finds placement in the framework of the researchs devoted to 
noise induced chaotic dynamics~\cite{huberman1980, gao1999, dennis2003},
however we are now dealing with a high dimensional system with 
a non trivial collective behavior. Furthermore, our approach,
despite being developed for a simple network model, can be easily
extended to a large class of complex networks.

The paper is organized as follows, Sect. II A is devoted to the introduction of the
DD model as well as of dynamical indicators able to characterize 
microscopic and macroscopic dynamics in this system.
In the same Subsection, the results concerning the dynamical evolution of
deterministic FC and diluted networks are briefly revisited.
In Sect. II B the stochastic model developed to 
mimic the dynamics of the diluted system is introduced. 
Three methods to estimate the Lyapunov spectrum in pulse-coupled
neural networks are revised in Sect. III. 
In the same Section the three methods are compared by applying them to
deterministic systems. Furthermore, the generalization 
of two of such methods to stochastic pulse-coupled networks with white and colored noise
is also presented. Sect. IV A deals with the analysis of the reconstructions of the  
microscopic and macroscopic features of the DD network via the stochastic approach. 
The Lyapunov analysis for the stochastic models is reported in Sect. IV B
and the results are compared with the ones obtained for the corresponding DD systems.
Finally, a summary and a brief discussion of the obtained results is reported in  Sect. V
together with a sketch of possible future developments.

\section{Models and Methods}

\subsection{Diluted Deterministic network}

\subsubsection{The Model}

We will focus our study on a diluted network of $N_D$ Leaky Integrate-and-Fire neurons (LIF), 
where the membrane potential $v_i$  of each neuron evolves according to
the following first order differential equation
\begin{equation}
\label{eq:dotV}
\dot{v}_i(t)= a - v_i(t) + g E_i(t) \qquad i = 1,\dots,N_D \quad ;
\end{equation}
where $a > 1$ represents a supra-threshold DC current and $g E_i$ the synaptic current, with $g > 0$ 
being the excitatory synaptic coupling.
Whenever the membrane potential of the $i$-th neuron reaches a fixed threshold $v_\Theta=1$, 
the neuron emits a pulse $p(t)$ transmitted, without any delay, to all the post-synaptic neurons and its
potential is reset to $v_R = 0$. In particular, the field $E_i(t)$  is given by the linear superposition of 
the pulses $p(t)$ received at the previous spike times $\{t_n\}$ by the $i$-th neuron from the pool of its 
pre-synaptic neurons. In this paper, in analogy with previous 
studies~\cite{Abbott1993Asynchronous,vanVreeswijk1996Partial,olmi2013coherent}, 
we assume that the transmitted pulse is an $\alpha$-function, namely $p(t) = \alpha^2 t \exp{\rm (- \alpha t)}$,
where $\alpha^{-1}$ is the width of the pulse.
In this case, the evolution of each field $E_i(t)$ is ruled by
the following second order differential equation
\begin{equation}
\label{eq:E_secondOrder}
\ddot E_i(t) +2\alpha\dot E_i(t)+\alpha^2 E_i(t)= 
\frac{\alpha^2}{K} \sum_{n|t_n < t} C_{j,i} \delta(t-t_n) \quad ;
\end{equation}
where $C_{j,i}$ is a $N_D\times N_D$ random matrix whose entries are $1$ if there is a synaptic 
connection from neuron $j$ to neuron $i$, and $0$ otherwise and $K$ is the number of pre-synaptic
connections of the $i$-th neuron. For a FC network $K=N$ and
all the fields are identical, since each neuron receives exactly 
the same sequence of spikes.
By introducing the auxiliary variable 
$P_i \equiv \alpha E_i +\dot{E}_i $, the second order differential equation \eqref{eq:E_secondOrder} 
can be rewritten as
\begin{equation}
\label{eq:E_and_P}
\dot{E_i} = P_i - \alpha E_i, \qquad \dot{P_i}=-\alpha P_i + \frac{\alpha^2}{K} 
\sum_{n|t_n < t} C_{j,i} \delta(t-t_n) \ .
\end{equation}

Therefore, the network evolution is ruled by the $3N_D$ Eqs. \eqref{eq:dotV} and \eqref{eq:E_and_P}
which can be exactly integrated between a spike event occuring at time $t_n$ and the successive one 
at time $t_{n+1}$, thus defining the following event driven map \cite{Zillmer2007,Olmi2010Oscillations}:
\begin{subequations}
\label{eq:diluted_map}
\begin{align}
\label{eq:E_map}
E_i(n+1)&=E_i(n) {\rm e}^{-\alpha \tau(n)}+P_i(n)\tau(n) 
{\rm e}^{-\alpha \tau(n)} \\
\label{qq}
P_i(n+1)&=P_i(n) {\rm e}^{-\alpha \tau(n)}+C_{m,i}\frac{\alpha^2}{K} \\
\label{V_map}
v_{i}(n+1)&=v_i(n){\rm e}^{-\tau(n)}+a(1-{\rm e}^{-\tau(n)})+g H_i(n) \, .
\end{align}
\end{subequations}
The $m$-th neuron is the next firing neuron, which
will reach the threshold at time $t_{n+1}$, i.e. $v_m(n+1) \equiv 1$.
One should notice that the event driven map is an exact rewriting
of the continuous time evolution of the system evaluated in correspondence of the spike 
emissions, therefore it can be considered as a Poincar\'e section of the original flux
in $3 N_D$ dimension. Indeed the event driven map is $3 N_D -1$ dimensional, since
the membrane potential of the firing neuron is always equal to one in correspondence
of the firing event. Here, $\tau(n)= t_{n+1}-t_n$ is the time between two consecutive spikes, which can be 
determined by solving the implicit transcendental equation
\begin{equation}
\label{eq:tau_implicit}
\tau(n)=\ln\left[\frac{a-v_m(n)}{a+gH_m(n)-1}\right] \, ;
\end{equation}
where the expression $H_i(n)$ appearing in equations (\ref{V_map}) and (\ref{eq:tau_implicit}) has the form
\begin{eqnarray}
\label{eq:H_i}
\nonumber H_i(n) &=& \frac{{\rm e}^{-\tau(n)} - {\rm e}^{-\alpha\tau(n)}}{\alpha-1} \left(E_i(n)+\frac{P_i(n)}{\alpha-1} \right) \\
 & &-\frac{\tau(n) {\rm e}^{-\alpha\tau(n)}}{(\alpha-1)} P_i(n) \, .
\end{eqnarray}

In this paper, we consider connectivity matrices $C_{j,i}$ corresponding to random graphs with directed
links and a fixed in-degree $K$ for each neuron~\cite{albert2002statistical}. This amounts to have a 
$\delta$-distribution centered at $K$ for the in-degrees, and a binomial distribution with average
$K$ for the out-degrees. In particular, we examine massively 
connected networks, where the in-degree grows proportionally to the system size,
namely our choice has been $K = 0.2 \times N_D$. As we have verified, the main results are not 
modified by considering Erd\"os-R\'{e}nyi distributions with an average in-degree equal to $K$.

\subsubsection{Microscopic and Macroscopic Dynamical Indicators}

In contrast to FC systems, the presence of dilution in the network induces fluctuations 
among the instantaneous values of the fields $\{E_i(t)\}$~\cite{Olmi2010Oscillations,Tattini2011Coherent}. 
These fluctuations can be estimated by evaluating the standard deviation $\sigma_E$ of the individual fields $E_i$ 
with respect to the ensemble average of the field $\bar{E}$, defined as follows
\begin{align}
\bar{E}(t) & = \frac{1}{N_D}\sum_{i=1}^{N_D} E_i(t) \\
\sigma_E(t) & = \left[\frac{1}{N_D} \sum_{i=1}^{N_D} d_i(t)^2 \right]^{1/2} \qquad ;
\end{align}
where $d_i(t) = E_i(t) - \bar{E}(t)$ denotes the instantaneous fluctuation of the $i$-th field
with respect to the ensemble average.
Similarly we can define $\bar{P}$ and $\sigma_P$. Obviously, for a FC network
$E_i \equiv \bar E$, $P_i \equiv \bar P$ and $\sigma_E = \sigma_P \equiv 0$.
In the following, we will consider an unconstrained time average of the fluctuations
$\langle \sigma_E \rangle$, as well as a conditional time average 
$\langle \sigma_E (\bar E, \bar P) \rangle$ evaluated whenever the 
value of the average fields falls within a box of dimension $\Delta \bar E \times \Delta \bar P$
centered at $(\bar E, \bar P)$.

To measure the level of correlation present in the field fluctuations $d_i(t)$,
we measure the associated autocorrelation function
\begin{equation}
\label{eq:ACF_definition}
C_E(\tau) = \displaystyle{\frac{\left\langle \frac{1}{N_D} \sum_{i=1}^{N_D}  d_i(t+\tau) d_i(t) \right\rangle }
{\left\langle (\sigma_E(t))^2 \right\rangle}} \; .
\end{equation}
where $\left\langle  \cdot \right\rangle$ indicates the average over time.
The time interval over which the fluctuations are correlated
can be estimated by measuring the decorrelation time $\tau_d$
from the the initial decay of $C_E(\tau)$.

The collective activity in the network can be studied by
examining the macroscopic attractor in the $(\bar E, \bar P)$-plane
as well as the distributions of the average fields $F(\bar{E})$ and $F(\bar{P})$. 
On the other hand, the microscopic dynamics has been characterized by considering the distribution 
$F(ISI)$ of the single neuron inter-spike intervals (ISIs) as well as 
the associated first return map.

\subsubsection{Diluted versus Fully Coupled Dynamics}

\begin{figure*}
\begin{center}
\includegraphics[width = 0.48\textwidth]{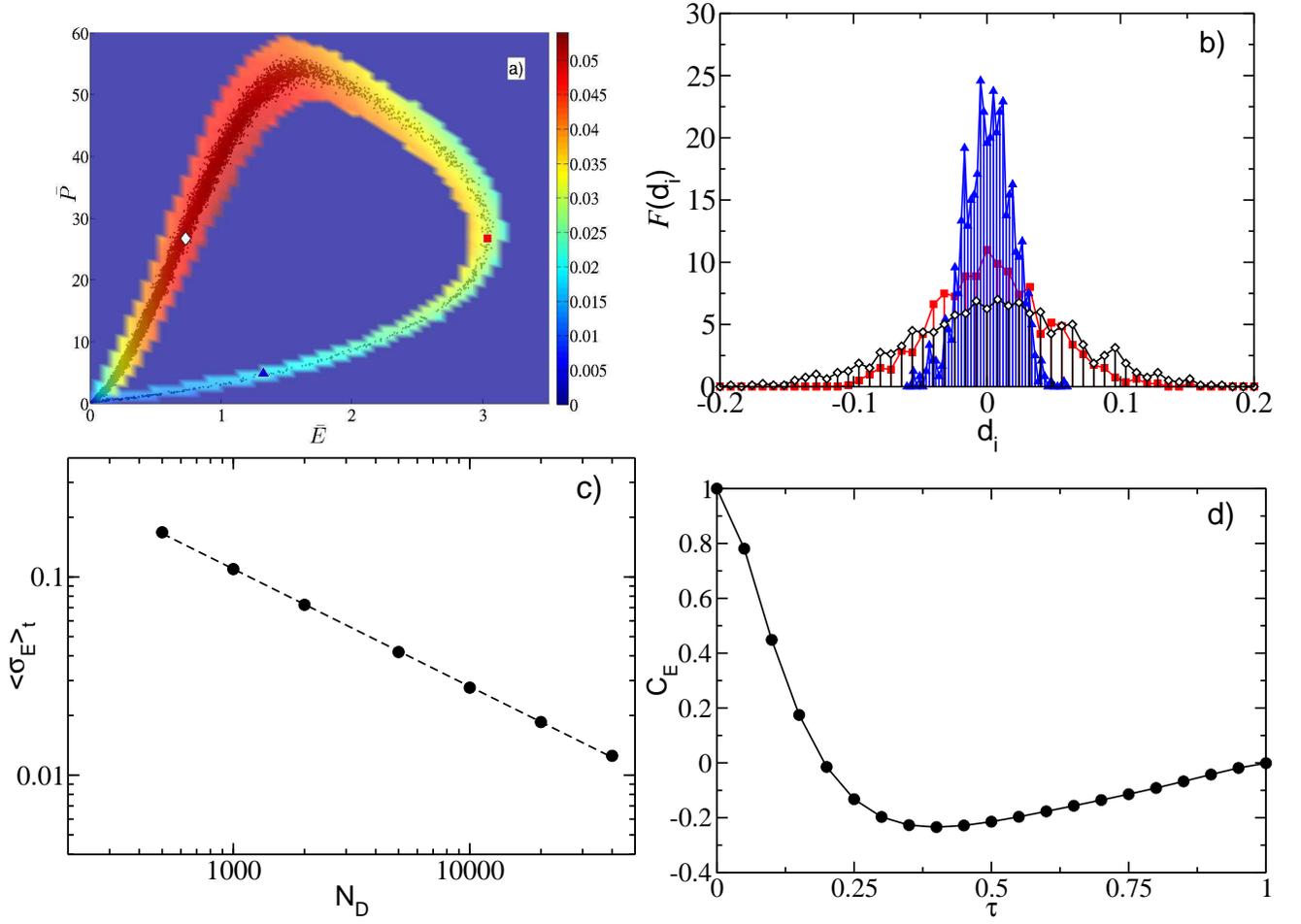}
\includegraphics[width = 0.48\textwidth]{FigsPaper/Fig1b.eps}
\includegraphics[width = 0.48\textwidth]{FigsPaper/Fig1c.eps}
\includegraphics[width = 0.48\textwidth]{FigsPaper/Fig1d.eps}
\end{center}
\caption{Characterization of the field fluctuations for a DD network. 
(a) Macroscopic attractor reported in the $(\bar E,\bar P))$ plane (black dots),
the colormap (superimposed on the attractor) quantifies the time averaged values of the fluctuations 
$<\sigma_E (\bar E, \bar P)>$. These are estimated over a grid $100\times 100$ 
with resolution $\Delta \bar E = 0.06$ and $\Delta \bar P = 0.8$ . (b) PDFs $F(d_i)$ of the deviation from the average 
field $d_i$ estimated in three different points along the attractor.
These points are indicated in panel (a) with the same symbol and color 
code. (c) Fluctuations of the fields $<\sigma_E>$ averaged both in time 
and all along the whole attractor as a function of the system size $N_D$
(filled circles). The dashed line indicates a power law fitting to the data,
namely $<\sigma_E> \propto N_D^{-0.6}$. (d) Autocorrelation function $C_E(\tau)$ of the 
fluctuations of the fields $d_i$. For all the reported data the parameters are 
fixed to $g=0.5$, $\alpha = 9$, $a = 1.05$ and $K = 0.2N_D$. The system size is set to
$N_D=500$, apart in panel c). The reported quantities have been 
evaluated over $10^5 - 10^6$ spikes, after discarding an initial transient 
of $10^6$ spikes.}
\label{fig:A9Charac}
\end{figure*}

As already mentioned in the Introduction, the dynamical regimes
observable for FC LIF networks, with post-synaptic potentials 
represented as $\alpha$-function, have
been analyzed in~\cite{Abbott1993Asynchronous,vanVreeswijk1996Partial}.
These regimes are the so-called splay state and partial synchronization.
The splay state is characterized by a constant value for the field $E$ and by a perfectly periodic evolution
of the single neurons, on the other hand in the partially synchronized regime, the common field reveals 
a perfectly periodic evolution, while the single neuron dynamics is quasi periodic~\cite{olmi2013coherent}.
In the present paper we will focus in the latter regime, where collective oscillations in the network activity 
are present, in this case the macroscopic attractor corresponds to a closed curve in the two dimensional 
$(\bar E, \bar P)$-plane. The introduction of random dilution in the system induces fluctuations $d_i$ in the 
fields $E_i$ with respect to their ensemble average $\bar E$. Therefore the collective attractor still 
resembles a closed orbit, but it has now a finite width whose value depends on the values 
of $(\bar E, \bar P)$ (see Fig.~\ref{fig:A9Charac} (a)).
As shown in Fig.~\ref{fig:A9Charac} (b), the fluctuations $d_i$ are approximately Gaussian distributed 
for any point $(\bar E, \bar P)$ along the curve.
Therefore, the $d_i$ can be characterized in terms of their standard deviation 
$\langle  \sigma_E \rangle$ averaged in time, this quantity, as previously shown in~\cite{Olmi2010Oscillations,Tattini2011Coherent},
vanishes in the thermodynamic limit for massively connected networks. 
Indeed this is verified also in the present case as shown
in Fig.~\ref{fig:A9Charac} (c), thus indicating that for sufficiently large system sizes 
one recovers the regular motion observed for FC systems. It should be recalled that
for sparse networks the fluctuations do not vanish, even for diverging system sizes~\cite{Tattini2011Coherent}.
Furthermore, the field fluctuations present a decorrelation time
$\tau_d \simeq 0.1$, measured from the decay of the autocorrelation function $C_E(\tau)$ (see Fig.~\ref{fig:A9Charac} (d)), 
which is essentially independent from the system size, as we have verified.

Another relevant aspect of the diluted system dynamics is that
the random dilution of the links renders the finite network chaotic.
In particular, for a massively connected network the system becomes regular in the thermodynamic limit,
while a sparse network remains chaotic even for $N_D \to \infty$~\cite{olmi2013coherent}.
This result suggests that the degree of chaoticity in the system is related
to the amplitude of the fluctuations $d_i(t)$ of the macroscopic fields.

\subsection{Fully Coupled Stochastic Network}

The question that we would like to address is whether the dynamics of the DD network
can be reproduced in terms of an equivalent FC network with additive stochastic terms. 
As a first approximation, we assume that the erratic dynamics of the DD system is essentially due
to the field fluctuations $d_i(t)$. Therefore, we rewrote the dynamics of the diluted system
as follows
\begin{subequations}
\label{eq:FullyCoup}
\begin{align}
\label{eq:dotE_FC}
\dot{E}(t)&= P(t)-\alpha E(t)  \\
\label{eq:dotP_FC}
\dot{P}(t)&= -\alpha P(t) + \frac{\alpha^2}{N_S} \sum_{n\mid t_n < t}\delta (t-t_n) \\
\label{eq:dotV_FC}
\dot{v}_i(t)&= a - v_i(t) + g E(t) + g \xi_i(t)  \quad i = 1,\dots,N_S 
\end{align}
\end{subequations}
where each neuron is driven by the same mean field term $E(t)$,
generated by the spikes emitted by all the neurons, plus an additive stochastic 
term  $\xi_i(t)$.  Notice also that we use a different number of neurons in the 
reduced model $N_S<N_D$ since the asymptotic evolution of a FC system is 
fairly well retrieved already with a relatively small number of neurons.
We will consider both white noise as well as colored one. In particular, for white Gaussian noise
\begin{equation}
\label{eq:whitenoise}
\left\langle \xi_{w_i} (t) \xi_{w_j} (t^\prime) \right\rangle = D^2 \delta_{i,j} \delta(t - t^\prime)
\end{equation}
with a zero average value, namely $\left\langle \xi_{w_i} \right\rangle = 0$.
For colored noise, we considered exponentially time correlated noise as
follows
\begin{equation}
\label{eq:OU}
\left\langle \xi_{OU_i} (t) \xi_{OU_j} (t^\prime) \right\rangle = \frac{D^2}{2\tau_d} \delta_{i,j} {\rm e^{-|t-t^\prime |/\tau_d}}
\end{equation}
where the average of the noise term is again zero. This is the so-called
Ornstein-Uhlenbeck (OU) noise, which can be obtained by integrating the following ordinary differential equation
\begin{equation}
\label{eq:OU_proc}
\dot{\xi}_{OU_i}(t)= -\frac{1}{\tau_d}\xi_{OU_i} + \frac{1}{\tau_d}\xi_{w_i}
\end{equation}
where $\xi_{w_i}$ is a Gaussian white noise, with the correlation defined in
Eq.~\eqref{eq:whitenoise}.

The main issue is to estimate the value of the noise amplitude $D$ to insert 
in Eqs. (\ref{eq:whitenoise}) and (\ref{eq:OU}) and of the correlation time 
of the OU process $\tau_d$ to obtain a good reconstruction of
the original dynamics. The latter parameter can be straightforwardly quantified from
the autocorrelation function decay, in particular we set $\tau_d = 0.1$. 
For the former quantity, as a first attempt, 
we set $D$ equal to the time averaged standard deviation of the fields $\langle \sigma_E \rangle$. However,
the quality of the reconstruction was not particularly good and 
this can be explained by the fact that the fluctuation amplitude is
state dependent, as shown in Fig.~\ref{fig:A9Charac} (b).
Therefore, we evaluated $\langle \sigma_E (\bar{E},\bar{P}) \rangle $ during simulations of
DD systems and we employed these quantities in the stochastic integration
of the FC system. In particular, we set $D =\langle \sigma_E ({E},{P}) \rangle $,
where $E$ and $P$ are now the values of the fields obtained during 
the simulation of the FC stochastic system.
   
We have performed the stochastic integration (see appendix \ref{app:stochasticInt} for details) 
by employing extremely small time steps.
Such a choice is not due requirements related to the 
the precision of the integration scheme (in particular in the white noise
case the integration would be exact, see \eqref{eq:white_prop}), 
but to the fact that for the evolution of
our system is crucial to detect the spike emissions with extremely accuracy.
Therefore, instead of recurring to more elaborate integration schemes~\cite{Mannella1999Absorbing},
we decided to use small integration time steps $h$ in order to accurately determine 
the threshold crossing of the membrane potentials even in presence of noise.

\section{Linear Stability Analysis}
\label{sec:linearStab}

We are not only interested in the reconstruction of the macroscopic and 
microscopic dynamical features of the DD system via the stochastic approach, 
but also in the reproduction of the linear stability properties of the original
model. The latter can be quantified in terms of the Lyapunov Spectrum $\{\lambda_i\}$,
which can be related to the average growth rates of infinitesimal volumes in the tangent space.
The Lyapunov spectrum has been estimated by considering the linearized evolution of the original
system and by applying the usual procedure developed by Benettin et al.~\cite{BenettinLyapunov1980}. 
Therefore, let us start from the formulation of the linearized evolution of the DD case by differentiating 
\eqref{eq:dotV} and \eqref{eq:E_and_P}, this reads as:
\begin{subequations}
\label{eq:tangDeterm}
\begin{align}
\delta \dot{E}_i &=  \delta E_i - \alpha \delta P_i   \quad \, \\
\delta \dot{P}_i &=  -\alpha \delta P_i \\
\delta \dot{v}_{i} &=  -\delta v_i + g \delta E_i\, \qquad i = 1,\dots,N_D \quad \quad;
\end{align}
\end{subequations}
where $\{\delta E_i, \delta P_i,\delta v_i \}$ is a $3 \times N_D$ vector in the tangent space.
In the following we will limit our analysis  to the maximal non zero Lyapunov exponent. 
 
It should be noted that two discontinuous events are present in the evolution of the original orbit:
namely, the spike emission, which affects the field variables $\{P_i\}$,
and the reset mechanism acting on the membrane potentials $\{v_i\}$.
However, these discontinuities are not explicitly present in the 
Ordinary Differential Equations (ODEs) representing
the tangent space evolution (\ref{eq:tangDeterm}). In the next sub-sections
we will report three different approaches on how to deal with these discontinuities
in deterministic systems, and the possible extension to Stochastic Differential Equations (SDEs) 
for two of them. The first approach requires the formulation of the dynamics in terms of 
an exact event driven map so we do not see the possibility to extend it to stochastic
systems. Instead, the other two methods concern the integration of ODEs with discontinuities
and they can be easily extended to SDEs. 

\subsection{Linearization of the Event Driven Map (LEDM)}

This approach can be applied
whenever it is possible to write the evolution between two
successive events in an exact manner, and the expression
(even implicit) of the time interval $\tau (n)$ between
two events is known. Here, we will focus on the method
introduced in~\cite{Zillmer2007} for networks of pulse coupled LIF
neurons. In this case it is possible to write explicitly
the linearization of the event driven map by differentiating
Eqs. (\ref{eq:diluted_map}),  \eqref{eq:tau_implicit} and \eqref{eq:H_i}.

The linearization reads as:
\begin{subequations}
\label{eq:tangEventDriven}
\begin{align}
\delta E_i(n+1) &=  {\rm e}^{-\alpha \tau(n)} \left[ \delta E_i(n) +\tau(n) \delta
P_i(n) \right] \nonumber
\\
 & - {\rm e}^{-\alpha \tau(n)} \left[\alpha E_i(n) + (\alpha \tau(n)-1) P_i(n) \right] \delta \tau(n)\,,\\
\delta P_i(n+1) &=  {\rm e}^{-\alpha \tau(n)} \left[ \delta P_i(n)-\alpha P_i(n) \delta \tau(n) \right]\, ,
\\
\delta v_{i}(n+1) &=  {\rm e}^{-\tau(n)} \left[ \delta v_i (n) + (a-v_i(n)) \delta \tau(n) \right] + g \delta H_i (n) 
\nonumber
\\
i & =1,\dots,N_D \quad ; \quad \delta v_m(n+1) \equiv 0 \, .
\end{align}
\end{subequations}
where $m$ is the index of the neuron firing at time $t_{n+1}$, 
while the condition $\delta v_m(n+1) \equiv 0$ is a consequence of the  
Poincar\'e section we are performing to derive the event driven map.

The evolution of the LEDM is completed by the expression for $\delta \tau(n)$,
which is 
\begin{eqnarray}
\delta \tau(n) =\tau_v \delta v_m(n) +\tau_E\delta E_m(n)+\tau_P\delta P_m(n)\ ,
\end{eqnarray}
where
\begin{eqnarray}
\tau_v:= \frac{\partial \tau}{\partial v_m}  \quad  , \quad
\tau_E:= \frac{\partial \tau}{\partial E_m}  \quad ,  \quad
\tau_P:= \frac{\partial \tau}{\partial P_m}  \quad .
\end{eqnarray}
More details regarding this method can be found in \cite{Olmi2010Oscillations} for a DD system.

\subsection{M\"uller-Dellago-Posch-Hoover (MDPH) Method}

A well known method used for the calculation of Lyapunov exponents for 
discontinuous flows has been introduced in~\cite{Lyap_Muller,dellago1996lyapunov} 
and it has been recently extended to integrate and 
fire neural models with refractory periods in~\cite{Lyap_Zhou} and
to piece-wise linear models of spiking neurons~\cite{coombes2012nonsmooth}.
Here we will present an application of this method to our DD neuronal model.
The approach consists of integrating in parallel the linearized
evolution \eqref{eq:tangDeterm} and the ODEs describing 
the evolution of the orbit, namely \eqref{eq:dotV} and \eqref{eq:E_and_P},
until one of the neurons reaches threshold. At this point the tangent
vector value should be updated, due to a discontinuous event, as explained below.
 
By following the notation used in~\cite{Lyap_Muller}, let us consider a dynamical system described by a flow equation
$\mathbf{\dot{x}} = \mathbf{f_1(x)}$ with a discontinuity defined by 
some implicit equation in terms of the state variables  $l\mathbf{(x)} = 0$.
The evolution at the discontinuity is defined in terms of a function
$\mathbf{g(x)}$ mapping the state of the system from 
the time immediately previous to the discontinuity to the one 
immediately after, i.e $\mathbf{x_+} = \mathbf{g(x_-)}$.
Finally, let us assume that the dynamics after the discontinuity
is ruled by a different flow equation, namely $\mathbf{\dot{x}} = 
\mathbf{f_2(x)}$. 

In this way the evolution in the real space is perfectly defined, while
the correction to the tangent space vector $\delta \mathbf{x}$, 
due to the discontinuity, can be expressed as follows
\begin{align}
\label{eq:dx+Muller}
\delta \mathbf{x_+} & = 
\boldsymbol{G(x_-)}\delta\boldsymbol{x_-} + \left[ 
\boldsymbol{G(x_-)f_1(x_-)}-\boldsymbol{f_2(x_+)} \right] \delta t ,
\end{align}

where, provided that specific solvability conditions are met~\cite{Lyap_Muller}

\begin{align}
\label{eq:dtMuller}\delta t & = - \frac{\boldsymbol{L(x_-)}\delta 
\boldsymbol{x_-}}{\boldsymbol{L(x_-)f_1(x_-)}} .
\end{align}
The notation $\mathbf{x_-}$ ($\mathbf{x_+}$) indicates the state of the 
system right in the moment $t^*$ of reaching the discontinuity (just after $t^*$).
Moreover, 
$$
\mathbf{L(x)} = \frac{\partial l\mathbf{(x)}}{\partial \mathbf{x}} 
\qquad \mathbf{G(x)} = \frac{\partial \mathbf{g(x)}}{\partial \mathbf{x}} \; .
$$

It is easy to show that for our DD system the flux is given by
$$ \mathbf{f_1(x)} = \mathbf{f_2(x)} = [ \mathbf{\dot E}, \mathbf{\dot P}, \mathbf{\dot v}] \quad,$$ 
and the map at the discontinuity reads as
\begin{eqnarray}
&& \mathbf{g(x)} = [E_{1_+},E_{2_+},...,P_{1_+},P_{2_+},...,v_{1_+},v_{2_+},...,v_{m_+},...] \nonumber\\
&&= [E_{1_-}, E_{2_-},..., P_{1_-} + \frac{\alpha ^2}{K}C_{m,1} , P_{2_-} + \frac{\alpha ^2}{K}C_{m,2},..., 
\nonumber \\
&& v_{1_+},v_{2_-},..., 0 ,...] \, ,
\end{eqnarray}
where  $m$ indicates the neuron firing at time $t^*$.
Furthermore, the firing condition can be expressed as the scalar function
$$l\mathbf{(x)} = v_m - 1 \, . $$ 

Therefore, a straightforward calculation gives us the corrections to perform in 
the tangent space to take into account the firing event at time $t^*$:
\begin{subequations}
\label{eq:tangCorrect_MUL}
\begin{align}
\delta E_{i_+} &=  \delta E_{i_-} - C_{m,i} \frac{\alpha ^2}{K} \delta t  \, , \\
\delta P_{i_+} &=  \delta P_{i_-} + C_{m,i} \frac{\alpha ^3}{K} \delta t \, , \\
\delta v_{m_+} &=  - \dot{v}_{m_+} \delta t   \, , \label{eq:deltaVcorr_MUL}
\end{align}
\end{subequations}
with
\begin{equation}
\label{dt}
\delta t = -  \frac{\delta v_{m_-}}{\dot{v}_{m_-}} \quad .
\end{equation}
Here, $\delta t$ is the (linear) correction to apply to the 
spike time of the reference orbit to obtain the firing
time of the perturbed trajectory.
This quantity can be evaluated from the linearization of 
the threshold condition $v_m = 1$, and this leads to the following expression
\begin{equation}
\label{eq:deltaT_Pol}
\delta t = - \frac{1}{\dot{v}_{m_-}} 
\left( \left. \frac{\partial v_{m}}{\partial E_m}\right|_- \delta E_{m_-}
+ \left. \frac{\partial v_{m}}{\partial P_m}\right|_- \delta P_{m_-} \right) \quad ,
\end{equation}
where all the quantities entering in the rhs of the above equation
are evaluated exactly at the spiking time.

\subsection{Olmi-Politi-Torcini (OPT) Method}

Recently, another approach has been proposed to deal with the discontinuities 
occurring in the context of pulse-coupled neural networks~\cite{olmi2014linear}.
In this context the dynamical evolution in the tangent space between two spike events
is ruled by the $3 \times N_D$ ODEs reported in \eqref{eq:tangDeterm}.
Whenever a spike is emitted in the network the tangent space vector components should be 
updated as follows:
\begin{subequations}
\label{eq:tangCorrect_POL}
\begin{align}
\delta E_{i_+} &= \delta E_{i_-} + \dot{E}_{i_-} {\delta t}  \, , \\
\delta P_{i_+} &= \delta P_{i_-} + \dot{P}_{i_-} \delta t  \, ,\\
\delta v_{i_+} &= \delta v_{i_-} + \dot{v}_{i_-} \delta t \, 
, \label{eq:deltaVcorr_POL}
\end{align}
\end{subequations}
where the expression for $\delta t$ is reported in Eq.~\eqref{dt}
and the corrective terms appearing in \eqref{eq:tangCorrect_POL} account
for the difference in the spiking times of the perturbed and unperturbed orbit.
It is clear that in this case, just after the firing event, the component of the 
tangent  vector corresponding  to the membrane potential of the firing neuron 
is exactly zero, i.e. $\delta v_{m_+} \equiv 0$. In this approach, 
the evolution in tangent space is still performed by taking into account the constraint due to the
Poincar\'e  section associated to the event driven map, meaning that this 
method is completely analogous to the LEDM.

\subsection{Comparison of the Different Methods}

In order to verify the agreement among the different approaches introduced above,
we perform numerical estimation of the maximal non zero Lyapunov exponent by employing
such methods for a FC deterministic network. In this case the system is
never chaotic and in particular we consider two situations where
the microscopic neuronal dynamics is either periodic or quasi-periodic.
The first regime corresponds to the so-called splay state (observable for $\alpha =3$ 
for the chosen parameters) and the latter one to the partially synchronized regime 
(observable for $\alpha = 9$). In both cases, it has been shown that the 
whole branch of the Lyapunov spectrum corresponding to the membrane potentials 
vanishes as $1/N^2$ in the thermodynamic limit~\cite{Olmi2010Oscillations}.
In order to test for the accuracy of the employed methods, we decided to consider finite 
size networks, with $N_D =50 - 200$, where the Lyapunov  exponents are extremely small.

It is important to remember that the definition of the LEDM and OPT methods
require a Poincar\'e section. Therefore, one degree of freedom, 
associated with the motion along the reference orbit, is removed from the dynamical
evolution and also the corresponding zero Lyapunov exponent from the Lyapunov spectrum. 
Conversely, the MDPH method is not based on a Poincar\'e section. This means that for a 
periodic motion the largest Lyapunov exponent, evaluated with LEDM and OPT methods, corresponds 
to the second Lyapunov exponent estimated with the MDPH. Similarly, when the neurons evolve 
quasi-periodically in time, the maximal non-zero Lyapunov exponent obtained with LEDM and OPT 
is the second one, while being the third one with the MDPH method. In summary, to test the accuracy 
of the algorithms  we compare in the periodic (quasi-periodic) case, the second (third) Lyapunov exponent as obtained by the MDPH method with the first (second) one obtained with the other two methods. We measured these exponents for different system sizes, namely $N_D =50$, 100 and 200. For all the 
considered parameter values and system sizes the agreement among the three methods is very good,
the discrepancies among the different estimations are always  of the order of $10^{-5} - 10^{-6}$,
as reported in Table~\ref{tab:Proof_fixed_time_lyap}.

We also tested the three algorithms for a diluted deterministic system where the maximal Lyapunov
exponent is definitely positive and its value is 2-3 orders of magnitude larger than the absolute
values of the Lyapunov exponents measured in the non chaotic situations. In this case to improve 
the precision of the integration scheme, we employed an event driven technique, where the
integration time step is variable and given by the solution of Eq.~\eqref{eq:tau_implicit}.
This implementation allows to avoid the interpolations required to find the firing times 
when the integration schemes with a fixed time step are used. 
Also for the DD systems the discrepancies among the three methods are of order $10^{-5} - 10^{-6}$
(as shown in Table~\ref{tab:Proof_fixed_time_lyap}), 
thus suggesting that these differences are most probably due to the slow convergence of the 
Lyapunov exponents to their asymptotic value rather than to the precision of the 
numerical integration. Nonetheless, these results confirm that the three approaches 
are essentially equivalent for the analysis of deterministic systems.

\begin{table*}[t]
\begin{tabular}{|| l || c c c | c || c c c | c ||} \toprule
& \multicolumn{4}{c ||}{$\alpha=3$, $g=0.4$, $a=1.3$, $K=N_D$} & \multicolumn{4}{c||}{$\alpha=9$, $g=0.4$, $a=1.3$, $K=N_D$}\\ 
\cline{2-5} \cline{6-9}
$N_D$ & LEDM &   OPT  & MDPH &   Max. Abs. Error&  LEDM & OPT & MDPH &   Max. Abs. Error\\ 
50  & -1.70$\times 10^{-4}$   & -1.67$\times 10^{-4}$ & -1.70$\times 10^{-4}$  &   2.00$\times 10^{-6}$ &   -1.83$\times 10^{-3}$ & -1.75$\times 10^{-3}$ & -1.76$\times 10^{-3}$ & 5.17$\times 10^{-5}$\\
100 & -4.25$\times 10^{-5}$   & -4.30$\times 10^{-5}$ & -4.38$\times 10^{-5}$  &   7.43$\times 10^{-7}$ &   -4.73$\times 10^{-4}$ & -4.60$\times 10^{-4}$ & -4.66$\times 10^{-4}$ & 6.87$\times 10^{-6}$\\
200 & -1.07$\times 10^{-5}$   & -1.14$\times 10^{-5}$ & -9.10$\times 10^{-6}$  &   1.29$\times 10^{-6}$ &   -1.19$\times 10^{-4}$ & -1.18$\times 10^{-4}$ &  -1.28$\times 10^{-4}$ & 5.87$\times 10^{-6}$\\
\botrule

\toprule
& \multicolumn{4}{c ||}{$\alpha=3$, $g = 0.5$, $a=1.05$, $K=0.2N_D$} & \multicolumn{4}{c||}{$\alpha=9$, $g = 0.5$, $a=1.05$, $K=0.2N_D$}\\ 
\cline{2-5} \cline{6-9}
$N_D$ & LEDM &   OPT  & MDPH &   Max. Abs. Error&  LEDM & OPT & MDPH &   Max. Abs. Error\\ 
200 & 9.4676$\times 10^{-3}$ & 9.4676$\times 10^{-3}$ & 9.4608$\times 10^{-3}$ & 4.50$\times 10^{-6}$ & 2.9515$\times 10^{-1}$ & 2.9515$\times 10^{-1}$   & 2.9514$\times 10^{-1}$ &  6.67$\times 10^{-6}$ \\
\botrule
\end{tabular}
\caption{Comparison of the maximal (non zero) Lyapunov exponents obtained with 
the three methods introduced in Sec.~\ref{sec:linearStab}, namely Linearization
of the Event Driven Map (LEDM), Olmi-Politi-Torcini (OPT) and 
M\"uller-Dellago-Posch-Hoover (MDPH) methods. 
Upper panel: For a deterministic FC network  in 
the periodic splay state regime (left set of parameters), and in the quasi periodic 
partially synchronized regime (right set of parameters). Lower panel: For a 
chaotic DD network in the asynchronous regime (left set of parameters) and 
in the partially synchronized regime (right set of parameters). In all cases, the system 
is first relaxed through a transient of $10^4$ spikes, after which the Lyapunov exponents 
are obtained by averaging over a period corresponding to $\simeq 10^7$ spike events. 
The reported errors are calculated as the maximal (absolute) difference between the average of 
the values obtained with the three methods and each single value. The MDPH and OPT 
estimates are obtained in the upper panel by integrating the system \eqref{eq:diluted_map} 
with a fixed time  step $h=5\times 10^{-6}$, while in the lower panel by employing
an event driven integration scheme, where the time step is variable and 
given by \eqref{eq:tau_implicit}.}
\label{tab:Proof_fixed_time_lyap}
\end{table*}

\subsection{Implementation for SDEs}

Let us explain in detail how we implement the evolution in the 
tangent space associated to the SDEs  Eqs.~\eqref{eq:FullyCoup}. 
For SDEs the estimation of the maximal Lyapunov exponent has been performed
by employing the MDPH and the OPT methods, since the LEDM cannot be used in the
case of a stochastic evolution, because it requires an exact knowledge
of the next firing time. For white additive noise, the linearized
equations for both methods have exactly the same form 
and they coincide with the expression in absence of noise reported in \eqref{eq:tangDeterm}. 
Notice that in this case we have a common field, therefore there are only two equations for 
the evolution of the infinitesimal perturbations $(\delta E, \delta P)$ of the field.
The stochastic nature of the process is reflected only in the evolution of the reference 
orbit around which the linearization is performed. The only {\it approximation} 
we have done in this case is the same adopted during the integration of the real space. 
Namely, at each firing time the values of the membrane potentials (entering in the
tangent space evolution) are simply evaluated as a linear interpolation between the values 
taken at the time step before and after the firing event and not by employing
some accurate stochastic propagator taking in account the presence 
of absorbing boundaries~\cite{Mannella1999Absorbing}.

In the case of OU noise the situation is more delicate, in particular
the equations for the evolution of the common field
correspond to Eqs. (\ref{eq:tangDeterm}a) and (\ref{eq:tangDeterm}b).
On the other hand the linearized equations for the membrane 
potentials and the OU noise terms now read as
\begin{subequations}
\label{eq:tangOU}
\begin{align}
\delta \dot{v}_{i} &=  -\delta v_i + g \delta E + g \delta \xi_{OU_i}  \\
\delta \dot{\xi}_{OU_i} &= -\frac{1}{\tau_d}\delta \xi_{OU_i} \qquad i = 1,\dots,N_D \quad .
\end{align}
\end{subequations}
It is easy to verify via \eqref{eq:dx+Muller} and \eqref{eq:dtMuller} that the
evolution of the Ornstein-Uhlenbeck process does not require extra corrections 
in correspondence of the firing events when the MDPH method is used, i.e $\delta \xi_{OU_{i+}} =  \delta \xi_{OU_{i-}}$. 

Instead, with the OPT approach each noise term $\delta \xi_{OU_i}$ should be updated
whenever a neuron spikes as follows
\begin{equation}
\label{eq:tangOU_corr}
\delta \xi_{OU_{i_+}} = \delta \xi_{OU_{i_-}} + \dot{\xi}_{OU_{i_-}}\delta t
\end{equation}
and $\delta t$ is now defined as 
\begin{eqnarray}
 &&\delta t = - \frac{\delta v_{m_-}}{\dot{v}_{m_-}} = \\
 && \frac{-1}{\dot{v}_{m_-}} 
\left( \left. \frac{\partial v_{m}}{\partial E}\right|_- \delta E_{-}
+ \left. \frac{\partial v_{m}}{\partial P}\right|_- \delta P_{-} + 
\left. \frac{\partial{v_m}}{\partial{\xi_{OU_{m}}}}\right|_- \delta \xi_{OU_{m_-}}\right)
\nonumber
\end{eqnarray}

\section{Results}

In this Section we examine the quality of the reconstruction 
of the macroscopic and microscopic features and of the stability properties
of the DD system in terms of SDEs representing a FC 
system subject to additive noise. In particular, we consider a massively
connected DD network with $K=0.2 N_D$ for various system sizes. namely
$500 \le N_D \le 10,000$. We reconstruct the
dynamics of these systems by employing a small FC stochastic system 
of size $N_S=100$, as we have verified that finite
size effects are of limited relevance for FC systems.
For each size of the DD system, we employ
as noise amplitude in the stochastic FC system
the standard deviation of the fluctuations of the corresponding DD fields.  
In particular, for the chosen
set-up (massively connected) as the system size
of the DD increases the amplitude of the fluctuations
of the fields decreases, vanishing in the thermodynamic limit
(as shown in Fig.\ref{fig:A9Charac} (c)).

\subsection{Macroscopic and Microscopic Dynamics}

\begin{figure*}
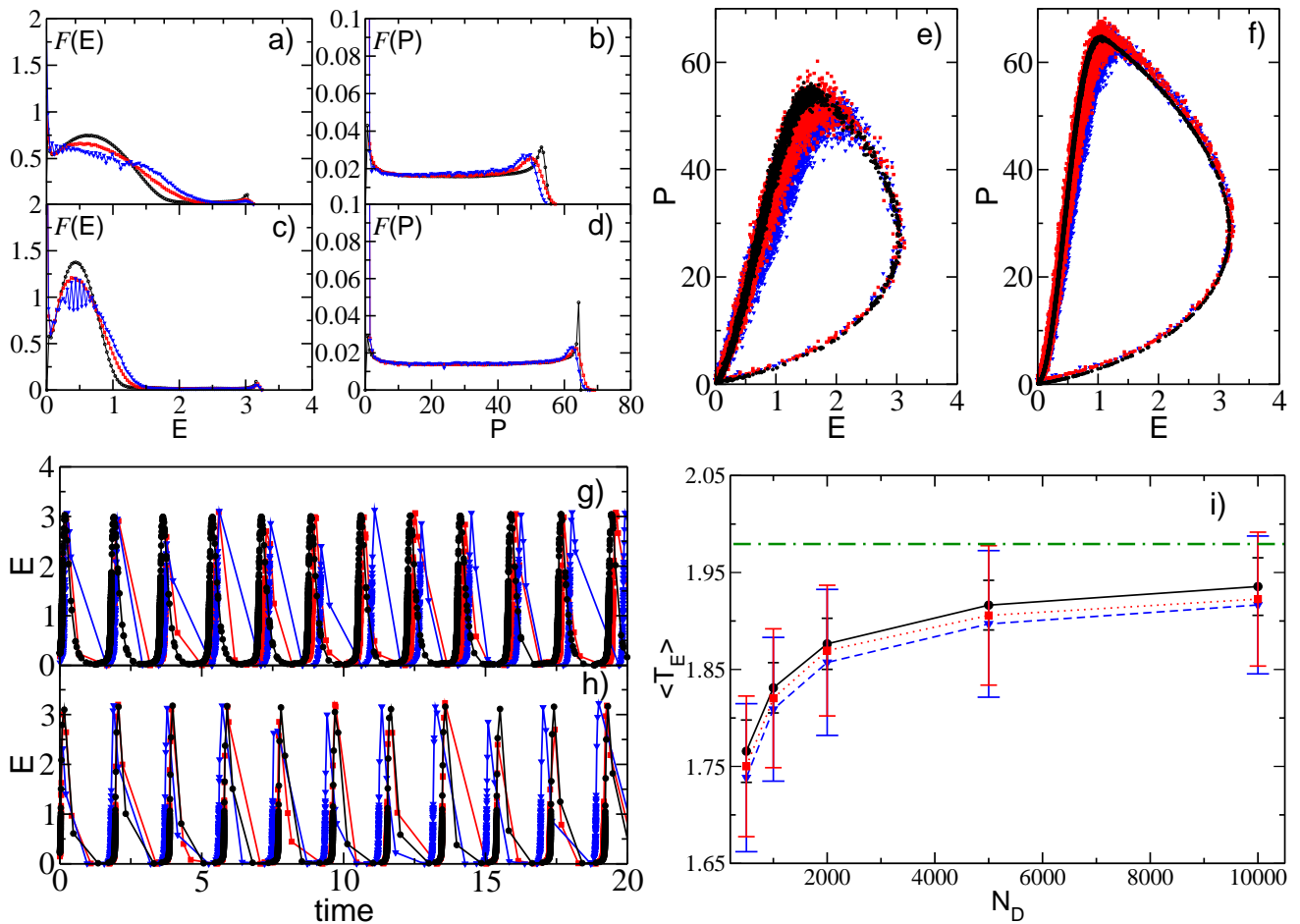

\begin{center}
\subfigure{\includegraphics[width = 0.48\textwidth]{FigsPaper/Fig2a-d.eps}}
\subfigure{\includegraphics[width = 0.48\textwidth]{FigsPaper/Fig2e-f.eps}}
\subfigure{\includegraphics[width = 0.48\textwidth]{FigsPaper/Fig2g-h.eps}}
\subfigure{\includegraphics[width = 0.48\textwidth]{FigsPaper/Fig2i.eps}}
\end{center}
\caption{Reconstruction of the macroscopic dynamics of the DD
system (black circle) in terms of white (blue triangles) and colored noise (red squares).
(a-d) Histograms of the macroscopic fields $E$ and $P$; (e-f) macroscopic attractors;
(g-h) time traces of the field $E$; (i) average period of the field $\langle T_E \rangle$
as a function of the system size $N_D$ of the DD system. Panels (a),(b),(e), and (g) refer
to $N_D = 500$, while panels (c), (d), (f), and (h) to $N_D = 5,000$.
In (g) and (h) the time traces have been shifted in order to ensure for the coincidence
of the time of occurrence of the first maximum in each trace. The periods reported in (i) have been obtained  
by measuring the time lapse between two consecutive maxima, the number of samples used for the 
calculation of $T_E$ is 5,000 data points,  and the (green) dot-dashed
line is the field period in a corresponding FC deterministic network equal to 1.98.
Reconstructed dynamics have been obtained with $N_S=100$ with an integration step 
$h = 5\times 10^{-6}$. Other parameters are as in Fig.~\ref{fig:A9Charac}.}
\label{fig:MacroscopicDyn}
\end{figure*}

In order to test for the quality of the reconstruction of the
macroscopic dynamics, we proceed to calculate 
the PDFs of the common field variables $E$ and $P$ in the FC set up, 
and compare them with the histograms of the average fields 
$\bar{E}$ and $\bar{P}$ as obtained in the DD case. 
These are reported in Fig. \ref{fig:MacroscopicDyn} (a-d) for two system
sizes of the diluted system, namely $N_D = 500$ and $N_D = 5,000$.
The agreement between the original PDFs and the 
reconstructed ones improve by passing from white to colored noise.
In particular, this is evident for the $F(E)$, since in the case
of white noise these distributions presents oscillations
which are absent in the original ones. 
The origin of these oscillations can be ascribed 
to the fact that the in presence of white noise
of equal amplitude along the whole macroscopic orbit
the field can be driven occasionally far from the original attractor.

When the colored noise is employed one observes
a better overall reconstruction of the macroscopic attractors
with respect to white noise. This is evident from
Fig. \ref{fig:MacroscopicDyn} (e-f),  the attractors obtained with OU noise
show less deviations from the DD attractor with respect to 
the white noise case, in particular around the maximal $\bar{P}$.
This is confirmed by considering the evolution
in time of the original and reconstructed fields.
The time traces of the fields  are compared in Fig.~\ref{fig:MacroscopicDyn} (g-h),
by matching the time occurrence of the first maximum of each field.
As one can see from the figure the OU reconstructed field follows 
reasonably well the original evolution, at least
in the considered time window, while the field of the system driven by
white noise shows, already after few oscillations period, 
a retard/advance with respect to the original one.
 
To render more quantitative this analysis, we have measured the average 
oscillation period of the field $\langle T_E \rangle$ for various sytem sizes $N_D$
of the DD networks and for the corresponding stochastic reconstructions with
white and OU noise. The results for all the considered system sizes are
displayed in Fig.~\ref{fig:MacroscopicDyn} (i). In the DD case $\langle T_E \rangle$
increases for increasing $N_D$ and tends towards the corresponding deterministic FC
value (dot-dashed line in the figure), this value will be reached in the thermodynamic limit, 
as expected~\cite{Olmi2010Oscillations}. Both the stochastic estimations slightly
underestimate the DD value, however while the periods obtained by employing OU noise exhibit
errors with respect to the DD values of the order $\simeq 0.4 - 0.9 \% $, 
the errors made with the white noise reconstruction are usually larger, 
namely between $1.0$ and $1.5 \% $.

Let us now examine the microscopic dynamics of the DD system,
this is  quite perculiar  for the chosen parameters,
corresponding to quasi-periodic evolution of the
membrane potentials of the  single neurons. Indeed, the single
neuron motion become exactly quasi-periodic only in 
the thermodinamic limit, where the regular FC dynamics
is recovered. For DD systems, as the ones here examined, 
the neurons evolve on an almost quasi-periodic orbit, apart
small chaotic fluctations. These motions can be analyzed by considering the 
inter-spike-interval (ISI) of the single cell, in particular we will
estimate the associated PDF $F(ISI)$ 
as well as the first return maps for the ISIs of the single neurons.

\begin{figure*}
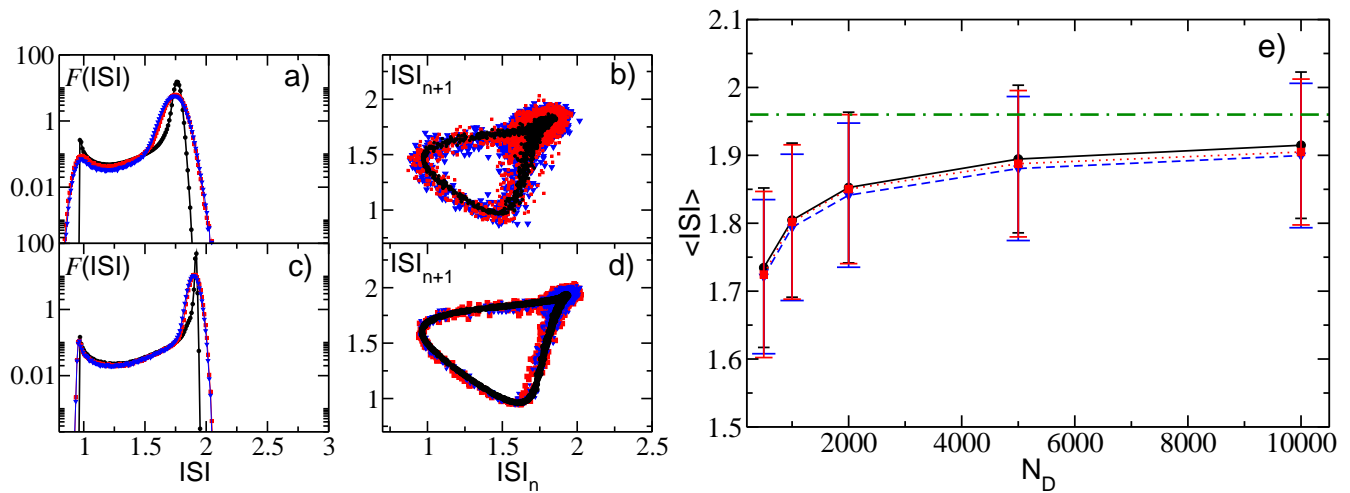

\centering
\subfigure{\includegraphics[width = 0.49\textwidth]{FigsPaper/Fig3a-d.eps}}
\subfigure{\includegraphics[width = 0.49\textwidth]{FigsPaper/Fig3e.eps}}
\caption{Reconstruction of the macroscopic dynamics of the DD
system (black circle) in terms of white (blue triangles) and colored noise (red squares).
(a) and (c) PDF $F(ISI)$ of the ISIs, (b) and (d) ISIs return maps.
 Panels (a) and (b) refer to $N_D = 500$, while panels (c) and (d) to $N_D = 5,000$.
(e) Average ISI  as a function of the DD system size $N_D$, 
the thick (green) dot-dashed line refers to the asymptotic value of the average ISI
in the corresponding FC deterministic set up, equal to 1.96 time units. 
The reconstructed dynamics are obtained with $N_S=100$ by employing an
integration step $h = 5\times 10^{-6}$. The other parameters are as in Fig.~\ref{fig:A9Charac}}
\label{fig:Return_Map}
\end{figure*}

The distributions $F(ISI)$ are reported in Figs.~\ref{fig:Return_Map}(a) and 
\ref{fig:Return_Map}(c)
for the same level of dilution and two different system sizes, namely
$N_D = 500$ and $N_D =5,000$. The $F(ISI)$ are defined over a finite range
of values, corresponding to the values taken by the ISIs during the neuron  
evolution, by increasing $N_D$, which corresponds to have smaller fluctuations
$\langle \sigma_E \rangle$, the $F(ISI)$ exhibit a sharper peak at large ISI 
and at the same time the return map appears to better approach
a closed line, as expected for quasi-periodic 
motions (see Figs.~\ref{fig:Return_Map}(b) and \ref{fig:Return_Map}(d)).

As far as the correpsonding SDEs are concerned,
the stochastic reconstruction is fairly good for the $F(ISI)$,
despite the fact that the distributions are now covering a slightly wider 
range with respect to the original PDFs.
The reconstructed return maps are {\it noisy} closed curves following closely 
the DD ones. By increasing $N_D$ the reconstruction improves
both with white and colored noise (as shown in Fig.~\ref{fig:Return_Map}(a-d)),
however it is difficult to distinguish among the two approaches relying
on these indicators. Therefore,  we have measured the average ISI
for different $N_D$ in the DD case and for the corresponding stochastic
dynamics. The results are reported in Fig.~\ref{fig:Return_Map}(e).
In the DD case, the $\langle ISI \rangle$ increases with $N_D$
approaching the FC deterministic limit (green dot dashed line in
the figure). The reconstructed $\langle ISI \rangle$ are slightly
under-estimating the deterministic results, however they reproduce 
quite closely the deterministic values.
From the figure it is clear that the OU reconstruction represents a better
approximation of the DD results for all the considered $N_D$, with errors ranging 
from $0.1$ to $0.6 \%$, with respect to the white noise results 
exhibiting discrepancies between $0.6 - 0.8 \%$ with respect to the DD values.
 
From the analysis of the macroscopic and microscopic features we
can conclude that the stochastic reconstruction improves by passing
from white to Ornstein-Uhlenbeck noise. This is particularly evident for 
the field $E$. The reason for this can be understood by considering the
evolution of the macroscopic field: $E$ displays a rapid rising phase of duration $\simeq 0.1 - 0.2$ followed by a relaxation period $\simeq 0.9 \times T_E$ (as shown in Figs.~\ref{fig:MacroscopicDyn}(g) 
and \ref{fig:MacroscopicDyn}(h)). Therefore, in order to properly reproduce this fast rise induced by the firing  of the most part of the neurons in the network, the time correlation of the fluctuations 
(on a time scale $\tau_d \simeq 0.1$) should be taken into account.

\subsection{Lyapunov Exponents}

\begin{figure}[t]
\includegraphics[width = 0.49\textwidth]{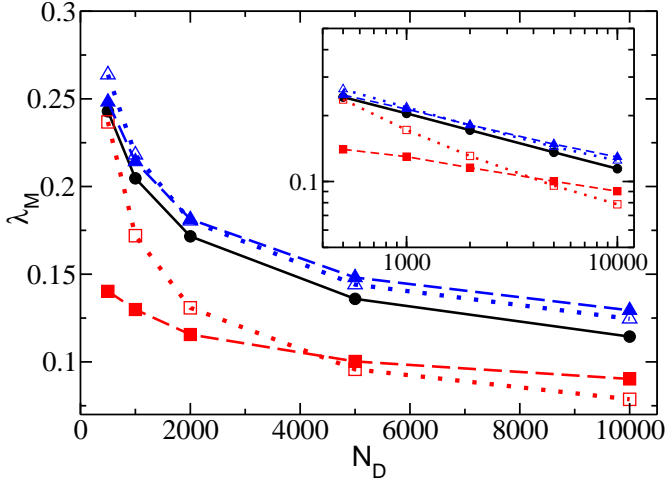}
\caption{Maximal Lyapunov exponent $\lambda_M$  as a function of the system size $N_D$
for the DD case (black circles and solid line) and the 
corresponding stochastic reconstructions evaluated with the MDPH method 
(red squares) and the OPT approach (blue triangles).
The stochastic results are reported for white (empty symbols and dotted lines)
and Ornstein-Uhlenbeck (filled symbols and dashed lines) noise.
Inset: plot in double logarithmic scale of $\lambda_M$ versus $N_D$. 
Reconstructed dynamics obtained with $N_S = 100$ and an integration step of $h=5\times 10^{-6}$.
In all cases, the system is relaxed during a transient of $10^6$ spikes and the Lyapunov
exponents are calculated by integrating  the tangent space for a period corresponding
to $10^7 - 10^8$ spikes. 
Other parameters as in Fig. \ref{fig:A9Charac}.}
\label{fig:Lyapunov}
\end{figure}

As already mentioned, the system is chaotic for the DD system
and the largest Lyapunov exponent vanishes in the thermodynamic limit
following a power law decay with $N_D$~\cite{Tattini2011Coherent}.
In particular, in the considered case we observe a decay 
$\lambda_M \propto N^{-\gamma}$ with $\gamma \simeq 0.25$ (as shown in the inset of Fig.~\ref{fig:Lyapunov}), 
which corresponds to a divergence of the maximal Lyapunov exponent with the averaged field fluctuations given by
$\lambda_M \propto \langle \sigma_E \rangle^{0.43}$.
On the other hand the FC deterministic
counter-part exhibits a perfectly regular dynamics for any
system size. Our aim is to reproduce the level of chaoticity 
present in the DD system by perturbing stochastically the FC system
with noise terms whose amplitude corresponds to that of the fluctuations of the deterministic 
fields $\{ E_i \}$, thus demonstrating that these fluctuations are at the origin of the chaotic behavior. 

The maximal Lyapunov exponents in the DD case have been estimated by employing the
LEDM method, while for the stochastic reconstructions we have used the
reconstructions. As it is evident from Fig.~\ref{fig:Lyapunov}, the MDPH
largely fails in reproducing the DD data, both for white and colored noise,
apart for the smallest system size here considered (namely, $N_D=500$) and white noise.
On the other hand, the OPT approach works quite well both with white and OU
noise over all the examined range of network sizes. The values obtained from the reconstructed
dynamics are always larger than the DD values, but while in the OU case the error 
in the estimation increases with $N_D$ and ranges from $2 \%$ at $N_D = 500$ to $13 \%$ at $N_D = 10,000$,
for the SDEs with white noise the error is of the order of $\simeq 5 - 9 \%$ and it seems not to depend 
on the considered system size. Furthermore, the OPT estimation of the maximal Lyapunov exponent 
is able to recover the correct power law scaling with $N_D$, in particular in the white (OU) noise case we 
have found an exponent $\gamma \simeq 0.25$ ($\gamma \simeq 0.22$). The exponent found for the white noise 
reconstruction coincides with the deterministic value.

A possible explanation for the failure of the MDPH method for the estimation of
the maximal Lyapunov exponent for a stochastic process with discontinuities relies on the definition
and implementation of the method. As shown in Eq.~(\ref{eq:tangCorrect_MUL}c) 
and Eq.~\eqref{dt} the {\it corrections} to be applied at each firing event depends only 
on the value of the derivative of the membrane potential of the firing neuron estimate just
before ($\dot v_{m_-}$) and after ($\dot v_{m_+}$) the event. These quantities 
depends on the actual value of the membrane potential at threshold and reset,
as a matter of fact we have assumed that these values are not affected by noise.
Maybe this assumption is too restrictive, 
however no better results have been obtained by the inclusion of stochastic terms. 
Instead, for the OPT approach the occurrence of a spike is taken into account by
modifying the values of the linearized variables in the tangent space on the basis of the 
time derivatives of the corresponding variables (in the real space) 
evaluated just before the spike emission (see Eq. \eqref{eq:tangCorrect_POL}). 
These time derivatives have been estimated as a linear 
interpolation between the values taken at the integration step immediately before and after the 
spike, therefore in their evaluation the stochastic evolution is somehow taken in account.

\section{Conclusions}

We have shown that the effect of the randomness in the distribution of the connections among neurons 
can be reproduced in terms of a perfectly regular (FC) network, where an additive noise term is introduced
in the evolution equations for the membrane potentials. Thus suggesting that noise or dilution can have
similar effects on the network dynamics (at least) in systems exhibiting collective oscillations.
These results opens new interesting directions for the study of the macroscopic activity of
large sparse (neural) networks, which can be mimicked in terms of few collective noisy variables. 
Furthermore, our analysis show that the stochastic approach is extremely convenient
from a computational point of view, since it allows to mimic the dynamics
of deterministic systems with $3 \times N_D$ variables by employing
$N_S + 2$ variables, where $N_S \simeq 100$ irrespectively 
of the size of the original network.

We have also discussed and critically re-examined the existing methods to
evaluate Lyapunov exponents for deterministic dynamical models with discontinuities,
and specifically for pulse-coupled systems. In particular, we have 
introduced a novel method to estimate stochastic
Lyapunov exponents for dynamical systems with discontinuities. Furthermore,
we have applied this novel approach in order to give a convincing evidence that the
fluctuations of the macroscopic variables acting on the membrane potentials are indeed
responsible for the presence of chaotic activity in diluted networks of LIF
excitatory neurons exhibiting collective oscillations. This is not obvious for any kind of LIF 
circuits, recent works~\cite{Zillmer2006,JhankeTimme2008PRL,MonteforteBalanced2012, AnguloArXiv} 
have shown the existence of linearly stable dynamics in sparse inhibitory networks where the fluctuations 
of the currents are responsible for the irregular activity of the system, in absence of chaotic motion.

The approach presented here appears to work reasonably well in presence of
collective oscillations in the macroscopic field (i.e. partial synchronization
in the network), while we have verified
that when the global activity is asynchronous the reconstructions
do not perform equally good. The origin of this discrepancy can be
traced back to the fact that the fluctuations of the fields are, 
in the asynchronous situation, almost periodic with decorrelation times 
$\cal{O}$($10^2$). Such slow decorrelation demands for more refined treatment 
of the noise term, like e.g. by considering harmonic noise terms~\cite{Shimansky1994Harmonic}.
Furthermore, a higher fidelity is needed in the tangent space
reconstruction since the maximal Lyapunov exponent is, in this case,
two orders of magnitude smaller than for the partially synchronized dynamics.

Our approach can be considered a sort of {\it stochastic mean field} 
version of the original system, in this regard it should be mentioned that in recent works,
the reconstruction of the dynamics of a diluted neural model quite similar to the
one analyzed here, has been successfully attempted by employing a deterministic
heterogeneous mean field (HMF) approach~\cite{di2014}. The HMF amounts to
introduce mean field variables associated to equivalence classes 
of neurons with the same in-degree, but it still maintains the heterogeneous
character of the diluted system, thus not allowing to clearly single out the source 
of the chaotic activity.

\acknowledgments

This work has been supported by the European Commission 
under the program ``Marie Curie Network for Initial Training", through the
project N. 289146, ``Neural Engineering Transformative Technologies (NETT)". D. 
A.-G. would like also to acknowledge
the partial support provided by ``Departamento Adminsitrativo de Ciencia 
Tecnologia e Innovacion - Colciencias" through
the program ``Doctorados en el exterior - 2013".  The authors would like to thank 
Prof. Andr\'e Longtin for very fruitful discussions and Dr. Mario Mulansky and Dr. Stefano Lepri 
for a careful reading of the manuscript prior to submission.

\appendix*
\section{Integration of the stochastic differential equations}
\label{app:stochasticInt}

Let us now examine how we can perform the integration of the SDEs (\ref{eq:FullyCoup}) 
for the white and OU noise. The integration of the ODEs for the 
fields $E$ and $P$ can be performed
without any approximation analogously to what done for the event driven map (\ref{eq:diluted_map}a) 
and (\ref{eq:diluted_map}b), since their evolution is completely deterministic. 
The integration of the equation
for the membrane potential (\ref{eq:FullyCoup}c) is instead performed in two steps,
first the deterministic part is integrated from time $t$ to $t+h$ as
\begin{equation}
\mathcal{F}_i(v) = v_i(t) {\rm e}^{-h}+a(1-{\rm e}^{-h})+ g H(h) \quad .
\end{equation}
Then the stochastic part is considered, for the white noise case, due to the linearity of 
the SDE the stochastic process can be integrated exactly~\cite{Mannella1999Absorbing},
and the solution reads as
\begin{align}
\label{eq:white_prop}
v_i(t+h) = \mathcal{F}_i(v)+D\sqrt{\frac{1}{2}(1-{\rm e}^{-2h})}\eta_i(t) \quad .
\end{align}
Here, the stochastic variable $\eta_i(t)$ is a spatio-temporal
uncorrelated random number, normally distributed  with zero average and unitary variance.
 
For the colored noise, instead, the integration of the SDE with accuracy 
${\cal O}[h^2]$ leads to the following set of equations~\cite{Toral2000Booklet}
\begin{align}
\label{eq:OU_prop}
\begin{array}{c l}
\xi_{OU_i}(0) & = \frac{D}{\sqrt{\tau_{d}}} \eta_i(0) \\
\xi_{OU_i}(t+h) & = \xi_{OU_i}(t) {\rm e}^{-h/\tau_{d}} + D\sqrt{\frac{1-{\rm e}^{-2h/\tau_{d}}}{\tau_{d}}} \eta_i(t)\\
v_i(t+h) & = \mathcal{F}_i(v) + h\xi_{OU_i}(t)
\end{array}
\end{align}
All the results reported throughout the paper should be interpreted in the Stratanovich sense.

The integration is performed with a constant time step $h \simeq 10^{-5}- 10^{-6}$.
In particular, we integrate exactly 
the equations for the field variables $E$ and $P$ for a time interval $h$, while the 
membrane potential is evaluated employing the stochastic propagators reported in 
Eqs. \eqref{eq:white_prop} or \eqref{eq:OU_prop}, depending whether we consider
white or OU noise. Whenever the membrane potential of one neuron 
overcomes threshold, we evaluate the crossing time $t^*$ and the values of all the
membrane potentials at $t^*$ via a linear interpolation. We then restart 
the integration with the values of the field variables and of the membrane potentials
evaluated at $t^*$, after resetting the potential of the neuron which has just fired.



\end{document}